\documentclass[aps,prd,amsmath,longbibliography]{revtex4-1}
\usepackage[margin=1in]{geometry}
\usepackage[utf8]{inputenc}
\usepackage{graphicx}
\usepackage{float}
\usepackage{indentfirst}

\begin{document}

\title{Superfast Quarks in Deuterium}
\author{Dmitriy N. Kim and Gerald A. Miller}
\affiliation{Department of Physics, University of Washington, Seattle, WA 98195-1560, USA}

\begin{abstract}
An extension to our previous study on Nuclear Parton Distribution Functions (nPDFs) using Light-Front Holographic Quantum Chromodynamics (LFHQCD) \cite{Kim:2022lng} is presented. We focus on applying the effects of nucleon motion inside the nucleus (Fermi motion/smearing) to deuterium, extending our nPDFs (and hence the DIS $F_2$ structure function for deuterium, $F_2^D$) to the \textit{superfast}, $x > 1$, region \cite{Frankfurt:1988nt}. We utilize four different deuteron wavefunctions (AV18, NijmI, NijmII, Nijm93) in this study. We find that our model, with no additional new parameters, is in excellent agreement with deuterium EMC ratio data obtained from the BONuS experiment \cite{Fenker:2008zz,CLAS:2011qvj,CLAS:2014jvt,Griffioen:2015hxa}. Looking beyond conventional nuclear physics, and in anticipation of ongoing 12 GeV experiments at Jefferson Lab, we use a LFHQCD ansatz to predict the contributions of an exotic six-quark state to $F_2^D$ in the superfast region. 
Our results are that the effects of using other potentials are about the same magnitude as six-quark effects \textemdash \, both have small effects in $x < 1$, but have significant contributions at $x > 1$. 
\end{abstract} 

\maketitle

\section{Introduction}
Parton distribution functions are ubiquitous in particle physics because they describe the relationship between Quantum Chromodynamics' (QCDs) basic degrees of freedom, quarks and gluons (partons), and the physical observable states, hadrons. This makes their nuclear counterparts, nuclear PDFs (nPDFs), indispensable tools towards understanding the emergence of intrinsic nuclear properties from the dynamics of their constituent partons. Programs to extract nPDFs have been, and are continued to be, supported with the operation of the Large Hadron Collider, the 12 GeV upgrade at Jefferson Lab (JLab), and upcoming Electron-Ion Collider (EIC). It is clear that understanding QCD interactions, how the dynamics of partons leads to the emergence of nucleons and nuclei, is of great interest and is a fundamental goal in nuclear science.\\

A domain where one can probe the interplay between nuclear physics and QCD's degrees of freedom lies in $x > 1$, dubbed the \textit{superfast} region \cite{Frankfurt:1988nt}. Superfast quarks are inherently generated by the nuclear environment. In other words, quarks with $x > 1$ cannot be produced by the QCD dynamics of a single free nucleon. Unlike quarks inside a free nucleon, a quark inside a nucleus can have a momentum fraction as large as $x = A$, where $A$ is the mass number of the nucleus. As such, investigations into superfast quarks are foundational towards uncovering intersections between nuclear physics and QCD. \\

Nuclear deep inelastic scattering (DIS) at high-$x$ gives us an opportunity to probe the superfast region. By extracting the DIS $F_2$ structure function of the nucleus, $F_2^A$, we can study the $x > 1$ momentum distributions of quarks inside a nucleus. To date, three experiments have undergone such an investigation: The BCDMS Collaboration at CERN \cite{BCDMS:1994ala}, the CCFR Collaboration at Fermi Lab \cite{CCFR:1999kbf}, and most recently at JLab \cite{Fomin:2010ei}. However, the trends of $F_2^A$ at high-$x$, extracted from all three experiments, do not agree with each other. The experiment at JLab originally reported results in agreement with BCDMS \cite{Fomin:2010ei}. However, a recent study by Freese et al. \cite{Freese:2015ebu} improved on the $Q^2$ evolution procedure used in the JLab study, obtaining results slightly in favor of BCDMS, but overall not strongly aligning with either of the two experiments. \\

It is clear that there is still much experimental work to be done in extracting the behavior of $F_2^A$ in the superfast region. The 12 GeV beam energy upgrade at JLab hopes to accomplish this by improving on its predecessors results. The experiment aims to extract $F_2^A$ at an even larger $x$ threshold, in kinematics where quasielastic contributions and scaling violations in the cross section are minimized \cite{arrington:2006pr, Hen:2014vua}. \\

In contrast to experimental progress, there has been little theoretical development in studying $F_2^A$ at $x > 1$, this is the goal of this paper. This study focuses on deuterium and the contents of this paper are as follows: In Sec. II, the light-cone convolution model developed in Refs. \cite{Frankfurt:1981mk,Frankfurt:1988nt} is introduced, which connects nuclear and bound-nucleon $F_2$ structure functions. The model achieves this by incorporating the conventional nuclear physics of Fermi motion (nucleon motion inside the nucleus) to bound-nucleon structure functions, outputting $F_2^A$ for $0 < x < A$ \textemdash \, our first steps into the superfast region. Following this, we will discuss what will be used as inputs into the convolution model: the light-cone density matrix of nucleons inside deuterium \cite{Frankfurt:1981mk,Frankfurt:1988nt}, and LFHQCD model for bound-nucleon PDFs \cite{Kim:2022lng}. Afterwards, results for the convolution model are presented for different deuteron wavefunctions (AV18, NijmI, NijmII, Nijm93). In Sec. III, a six-quark LFHQCD ansatz for deuterium is developed and incorporated into results in Sec. II. Our concluding remarks/discussion are given in Sec. IV.


\section{Nuclear Structure Functions}
The central theoretical objective to be addressed is the nuclear structure function $F_2^A(x,Q^2)$. Within the parton model, $F_2^A(x,Q^2)$ is connected to unpolarized nPDFs of flavor $f$, $f^{A}(x,Q^2)$. PDFs are not directly measurable, but are extracted by fitting phenomenological parameterizations to measured cross sections. Worldwide scientific programs, pioneered by the HERA experiment at CERN \cite{H1:2009pze}, have resulted in reliable proton PDFs. However, although parametrizations of nPDFs exist for several nuclei (e.g. \cite{Eskola:2016oht,Kovarik:2015cma}), they are not as robust due to a lack of experimental data over a wide kinematic range. With this in mind, developing a theoretical relationship between nPDFs and nucleonic PDFs is an efficient way to study $F_2^A(x,Q^2)$ in the absence of dependable nPDFs. This can be achieved by expressing nPDFs as a convolution between nucleonic PDFs and a nuclear light-cone density matrix, incorporating the effects of Fermi motion to nucleonic PDFs (see Refs. \cite{Frankfurt:1981mk,Frankfurt:1988nt} for discussion and derivation). Furthermore, discovery of the EMC effect (see original work \cite{EuropeanMuon:1983wih}) tells us that the momentum distributions of quarks inside bound nucleons are different than those of free nucleons. Therefore, nucleonic PDFs in the convolution model should be replaced with respective \textit{bound} PDFs. Thus for $F_2^A(x,Q^2)$, in the Bjorken limit (photon virtuality, $Q^2$, and energy, $\nu$, go to infinity at fixed $x = Q^2 / 2 m \nu$. Note that in this study, we are using isospin symmetry where $m = (m_p + m_n) / 2$, introducing negligible errors in our calculations), the convolution formula takes the following form \cite{Frankfurt:1981mk,Frankfurt:1988nt}:

\begin{equation}\label{convolution_formula}
    F_2^A(x,Q^2) = \sum_N\int_{x}^{A} \frac{d\alpha}{\alpha} \int d^2\boldsymbol{k}_\perp \,
    \rho_{N/A}(\alpha,\boldsymbol{k}_\perp) \, 
    \tilde{F}_2^N(x/\alpha, \alpha, \boldsymbol{k}_{\perp}, Q^2).
\end{equation}

\noindent Where $\rho_{N/A}(\alpha,\boldsymbol{k}_\perp)$ is the light-cone density matrix of nucleon $N$ inside nucleus $A$, $\tilde{F}_2^N(x/\alpha, \alpha, \boldsymbol{k}_{\perp}, Q^2)$ is the bound nucleon $F_2$ structure function, $\boldsymbol{k}_{\perp}$ is the transverse momentum of the nucleon, and $\alpha = A \, k^+ / k^+_A$ is the scaled light cone momentum fraction carried by nucleon $N$ inside nucleus $A$. Note that $\tilde{F}_2^N$ is a function of $\alpha$ and $\boldsymbol{k}_{\perp}$, in addition to $x/ \alpha$ and $Q^2$, due to medium modifications. \\

In this study, we investigate the $F_2$ structure function of the deuteron, $F_2^D(x,Q^2)$. To do so, Eq. (\ref{convolution_formula}) tells us that we need to determine its light-cone density matrix, $\rho_{N/D}(\alpha,\boldsymbol{k}_\perp)$, and bound nucleon structure functions, $\tilde{F}_2^N(x/\alpha, \alpha, \boldsymbol{k}_{\perp}, Q^2)$.

\subsection{Deuteron Light-cone Density Matrix}
The nuclear light-cone density matrix for nucleon $N$ inside nucleus $A$ is formally defined in terms of the nuclear wavefunction, $\psi_A$, \cite{Frankfurt:1981mk,Frankfurt:1988nt}:

\begin{equation}\label{lightcone_density_matrix}
\begin{split}
    \rho_{N/A}(\alpha,
    \boldsymbol{k}_\perp) & = \int \left[\prod_{i = 1}^A \frac{d\alpha_i}{\alpha_i} d^2\boldsymbol{k}_{i\perp} \right] \, \psi^{\dagger}_A(\alpha_1, ..., \alpha_A,\boldsymbol{k}_{1\perp}, ..., \boldsymbol{k}_{A\perp}) \, \psi_A(\alpha_1, ..., \alpha_A,\boldsymbol{k}_{1\perp}, ..., \boldsymbol{k}_{A\perp}) \\
    & \quad \quad \times \delta^{(1)} \left(1 - \frac{\sum_{i = 1}^A \alpha_i}{A} \right) \, \delta^{(2)} \left( \sum_{i = 1}^A \boldsymbol{k}_{i\perp} \right) \left\{\sum_{i = 1}^{(Z, \, A - Z)} \alpha_i \, \delta^{(1)}(\alpha - \alpha_i)\, \delta^{(2)}(\boldsymbol{k}_{\perp} - \boldsymbol{k}_{i\perp}) \right\},
\end{split}
\end{equation}

\noindent where the upper limit of the sum in the curly brackets is $Z$ for the proton and $A - Z$ for the neutron. The light-cone density matrix obeys the baryon and momentum sum rules:

\begin{equation}\label{baryon_sum_rule}
    \sum_N \int_0^A \frac{d \alpha}{\alpha} \int d^2 \boldsymbol{k}_{\perp} \, \rho_{N/A}(\alpha, \boldsymbol{k}_{\perp}) = A,
\end{equation}

\begin{equation}\label{momentum_sum_rule}
    \sum_N \int_0^A \frac{d \alpha}{\alpha} \int d^2 \boldsymbol{k}_{\perp} \, \alpha \, \rho_{N/A}(\alpha, \boldsymbol{k}_{\perp}) = A.
\end{equation}

Refs. \cite{Frankfurt:1981mk,Frankfurt:1988nt}, by neglecting all but the nucelonic degrees of freedom and identifying the internal $pn$ configurations in the deuteron through the $pn$ light-cone momentum,

\begin{equation}\label{pn_momentum}
    k = \sqrt{\frac{m^2 + k_{\perp}^2}{\alpha (2 - \alpha)} - m^2},
\end{equation}

\noindent are able to connect the deuteron light-cone wavefunction, $\psi_D$, to its non-relativistic wavefunction, $\psi_{NR}$: 

\begin{equation}\label{wavefunction_realtion}
    |\psi_D(k)|^2 = |\psi_{NR}(k)|^2 \sqrt{m^2 + k^2}.
\end{equation}

\noindent Eq. (\ref{wavefunction_realtion}) leads to the following expression for the light-cone density matrix of protons and neutrons in deuterium:

\begin{equation}\label{deut_lc_density}
    \rho_{pn/D}(\alpha, \textbf{k}_{\perp}) = \frac{|\psi_{NR}(k)|^2}{2 - \alpha} \sqrt{m^2 + k^2},
\end{equation}

\noindent where the subscript $pn$ was used as the proton and neutron have identical light-cone density maticies in the deuteron. One can check that Eq. (\ref{deut_lc_density}) obeys both sum rules. 

\subsection{Bound Nucleon $F_2$ Structure Function}
For the bound nucleon $F_2$ structure function, we utilize the phenomenological LFHQCD-based model from Ref. \cite{Kim:2022lng} \textemdash \, which we will refer as the nuclear LFHQCD (nLFHQCD) model for brevity. The model outputs closed-form expressions for valence nPDFs that are dependent on two phenomenological parameters, $\delta r_{p/A}$ and $\delta r_{n/A}$. The parameters are proportional to the average virtuality, a quantity that measures the average off-shellness of a nucleon is inside the nucleus. The nPDFs were used to construct $\tilde{F}_2^N$, given by:

\begin{equation}\label{lfhqcd_mod_F_2}
    \tilde{F}_2^N(x,Q^2_o) = \frac{4}{9} \, x \, \tilde{u}^N(x,Q^2_o) + \frac{1}{9} \, x \, \tilde{d}^N(x,Q^2_o) = F_2^N(x,Q^2_o) + \frac{5}{3} \, x \, (q_4(x,Q^2_o) - q_3(x,Q^2_o) ) \, \delta r_{N/A}, 
\end{equation}

\begin{equation} \label{mod_p_u}
    \tilde{u}^p(x,Q^2_o) = \left(\frac{3}{2} - 3\delta r_{p/A}\right)q_3(x,Q^2_o) + \left(\frac{1}{2} + 3\delta r_{p/A}\right)q_4(x,Q^2_o),
\end{equation}

\begin{equation} \label{mod_p_d}
    \tilde{d}^p(x,Q^2_o) = \left( - 3\delta  r_{p/A}\right)q_3(x,Q^2_o) + \left(1 + 3\delta r_{p/A}\right)q_4(x,Q^2_o).
\end{equation}

\noindent Where $\tilde{u}^p(x)$ and $\tilde{d}^p(x)$ are the medium-modified proton up and down valence PDFs respectively, $F_2^N$ is the structure function of free nucleon $N$, and $Q_o$ is the matching scale between LFHQCD and pQCD, $Q_0=1.06 \pm 0.15 $ GeV \cite{deTeramond:2018ecg}. The neutron valence PDFs are obtained by replacing $\tilde{u}^p \rightarrow \tilde{d}^n$, $\tilde{d}^p \rightarrow \tilde{u}^n$, and $\delta r_{p/A} \rightarrow \delta r_{n/A}$. The function $q_\tau(x,Q^2_o)$ is given by:

\begin{equation} \label{q_tau}
    q_\tau(x,Q^2_o) = \frac{\Gamma \left(\tau -\frac{1}{2}\right)}{\sqrt{\pi } \Gamma (\tau -1)}\big(1- w(x)\big)^{\tau-2}\, w(x)^{-1/2}\, w'(x),
\end{equation}

\begin{equation}\label{w_x}
    w(x) = x^{1-x} e^{-a (1-x)^2},
\end{equation}

\noindent with normalization

\begin{equation}
    \int_0^1 dx \, q_{\tau}(x,Q_o^2) = 1.
\end{equation}

\noindent Where the flavor-independent parameter $a = 0.531 \pm 0.037$ and $\tau$ us the number of constituents. The normalization of $q_{\tau}(x,Q_o^2)$ is  The modified structure functions were used to construct $F_2^A$, and numerical results for the phenomenological parameters were obtained by fitting to EMC ratio data in the region $0.3 \leq x \leq 0.7$ (EMC region). The results of the fitting were successful in reproducing EMC ratio data for a variety of nuclei. Having the correct behavior in the EMC region, the nLFHQCD model is an excellent candidate for $\tilde{F}_2^N$.\\

In the nLFHQCD model, the virtuality is not kinematical. In other words, nLFHQCD uses the average virtuality of a bound nucleon inside a nucleus, characterized by a single, nucleus-dependent number, and does not depend on detailed values of nucleon kinematics. This is because the model treats the nuclear potential as a constant. To improve on this we will use an ansatz motivated by results in Ref. \cite{Kim:2022lng}, where $\delta r_{p/A}(\alpha,\boldsymbol{k}_{\perp})$ and $\delta r_{n/A}(\alpha,\boldsymbol{k}_{\perp})$ are proportional to virtuality which depends on nucleon kinematics, up to a constant:

\begin{equation} \label{delta_r_ansatz}
    \delta r_{N/A}(\alpha,\boldsymbol{k}_{\perp}) = - \eta \, \mathcal{V}_{N/A}(\alpha, \boldsymbol{k}_{\perp}) \, \theta(1 - \alpha).
\end{equation}

\noindent Here $\mathcal{V}_{N/A}(\alpha, \boldsymbol{k}_{\perp})$ is the virtuality of bound nucleon $N$ inside nucleus $A$, $\eta$ is a nucleus-independent fitting parameter, and $\theta(1-\alpha)$ is the Heaviside step function. The nLFHQCD model was constructed to incorporate off-shell effects for single nucleons that are limited to carry $\alpha < 1$ momentum fraction. Therefore, extending off-shell effects beyond $\alpha > 1$ would exceeding the model's limits of applicability, hence the Heaviside step function in Eq. (\ref{delta_r_ansatz}). 

For the deuteron, in the center of mass frame, we use the following definition for virtuality:

\begin{equation}\label{virtuality}
    \mathcal{V}_{pn/D}(\alpha,\boldsymbol{k}_{\perp}) \equiv \frac{k_D^- - (k_p^- + k_n^-)}{k_D^-} = \frac{1}{m_D^2} \left( m_D^2 - 4\frac{m^2 + k_{\perp}^2}{\alpha (2 - \alpha)} \right),
\end{equation}

\noindent instead of $\mathcal{V} \equiv (k^2 - m^2) / m^2$, as used in Ref. \cite{Kim:2022lng}. This is because the convolution formula in Eq. (\ref{convolution_formula}) was obtained by using light-cone perturbation theory, where intermediate states are on mass-shell, $k^2 = m^2$, but off their energy shells, $k_D^- \neq k_p^- + k_n^-$. Eq. (\ref{virtuality}) is constructed to be negative, which is required to obtain the correct modification for the EMC region, $0.3 < x < 0.7$, in the nLFHQCD model. The connection between 'light-cone' virtuality, Eq. (\ref{virtuality}), and 'Feynman' virtuality, $\mathcal{V} \equiv (k^2 - m^2) / m^2$, within the BLC-PLC model used in Ref. \cite{Kim:2022lng} is discussed in Appendix A.\\

\section{Light-Cone Convolution Model Results}
For the nucleon momentum distribution in deuterium, $|\psi_{NR}(k)|^2$, we used the AV18 wavefunction \cite{Wiringa:2013ala}, unless stated otherwise, and the following normalization was used:

\begin{equation} \label{normalization}
    \int_0^\infty |\psi_{NR}(k)|^2 4 \pi k^2 dk = 1
\end{equation}

\begin{figure}[H]
    \centering
    \includegraphics[width=6in]{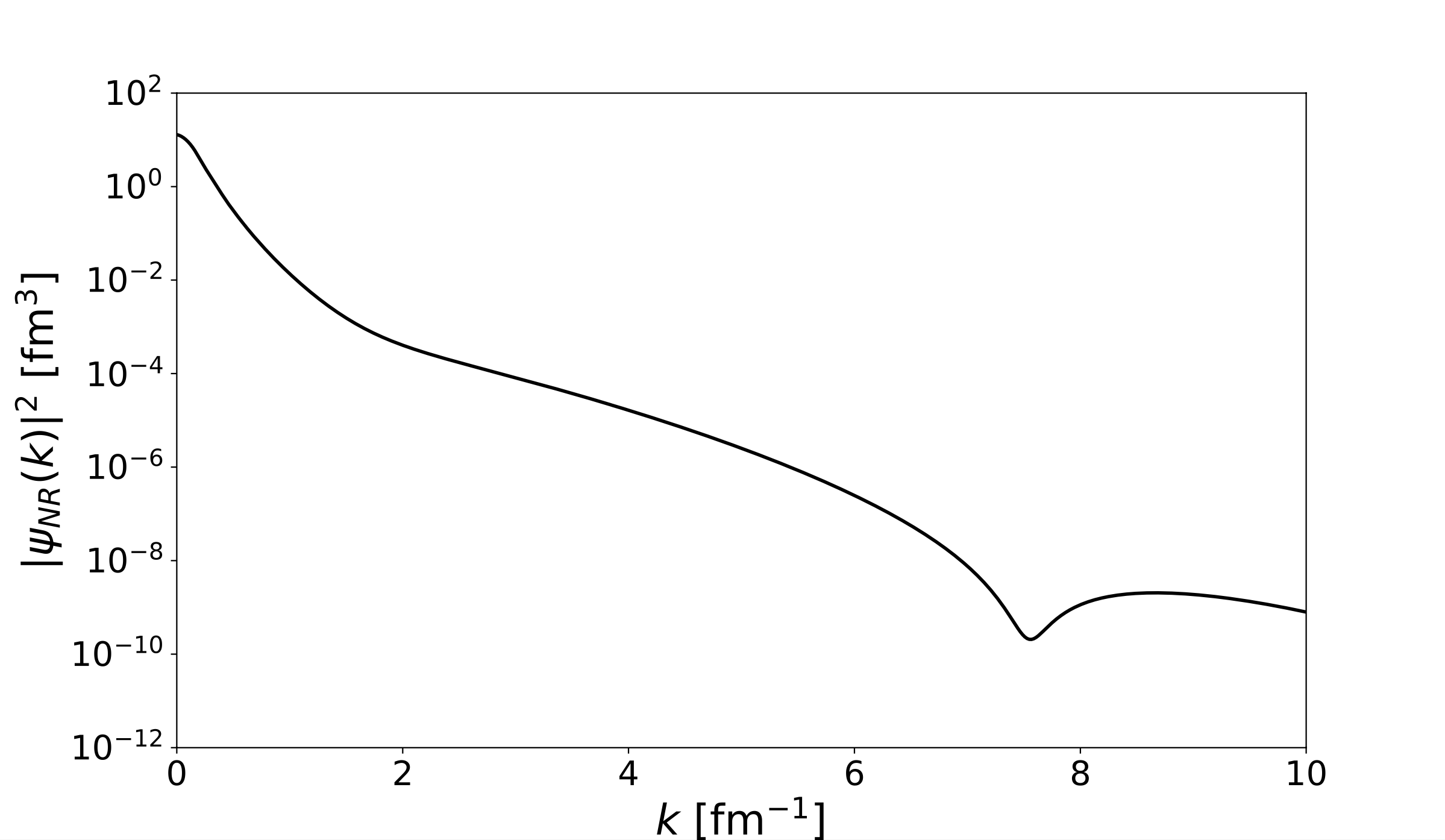}
    \caption{A plot of the interpolated AV18 nucleon momentum distribution in deuterium in the region $0 < k < 10$ fm$^{-1}$. }
    \label{fig:av18_dist}
\end{figure}

\noindent The momentum distribution was tabulated for finite points of $k$, the magnitude of the total momentum, and an interpolator was used. Fig. \ref{fig:av18_dist} presents the nucleon momentum distribution in deuterium for a $k$-range of $0 < k < 10$ fm$^{-1}$. The tabulated data was in the range $0 < k < 20$ fm$^{-1}$, anything outside this range was evaluated to be 0.

\begin{figure}[H]
    \centering
    \includegraphics[width = 6in]{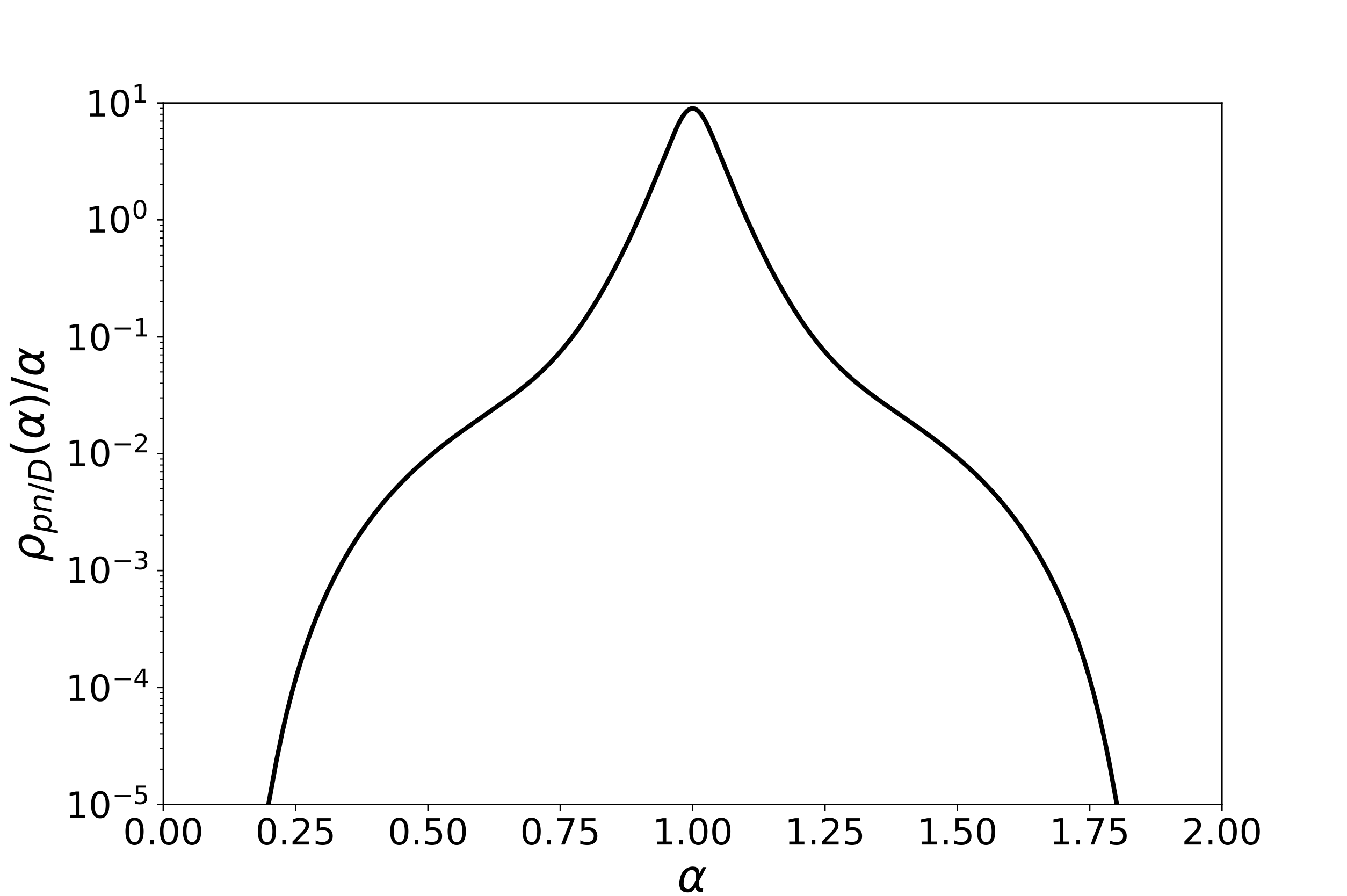}
    \caption{The proton and neutron light-cone density matricies for deuterium, divided by $\alpha$, as a function of $\alpha$, the momentum fraction of the nucleon in the nucleus weighted by $A$.}
    \label{fig:deut_lf_dist}
\end{figure}

Using the AV18 momentum distribution, we obtained the light-cone density matrix by using Eq. (\ref{deut_lc_density}). The light-cone density matrix for nucleons in deuterium, divided by $\alpha$, is presented in Fig. \ref{fig:deut_lf_dist}.

\begin{figure}[H]
    \centering
    \includegraphics[width = 6.5in]{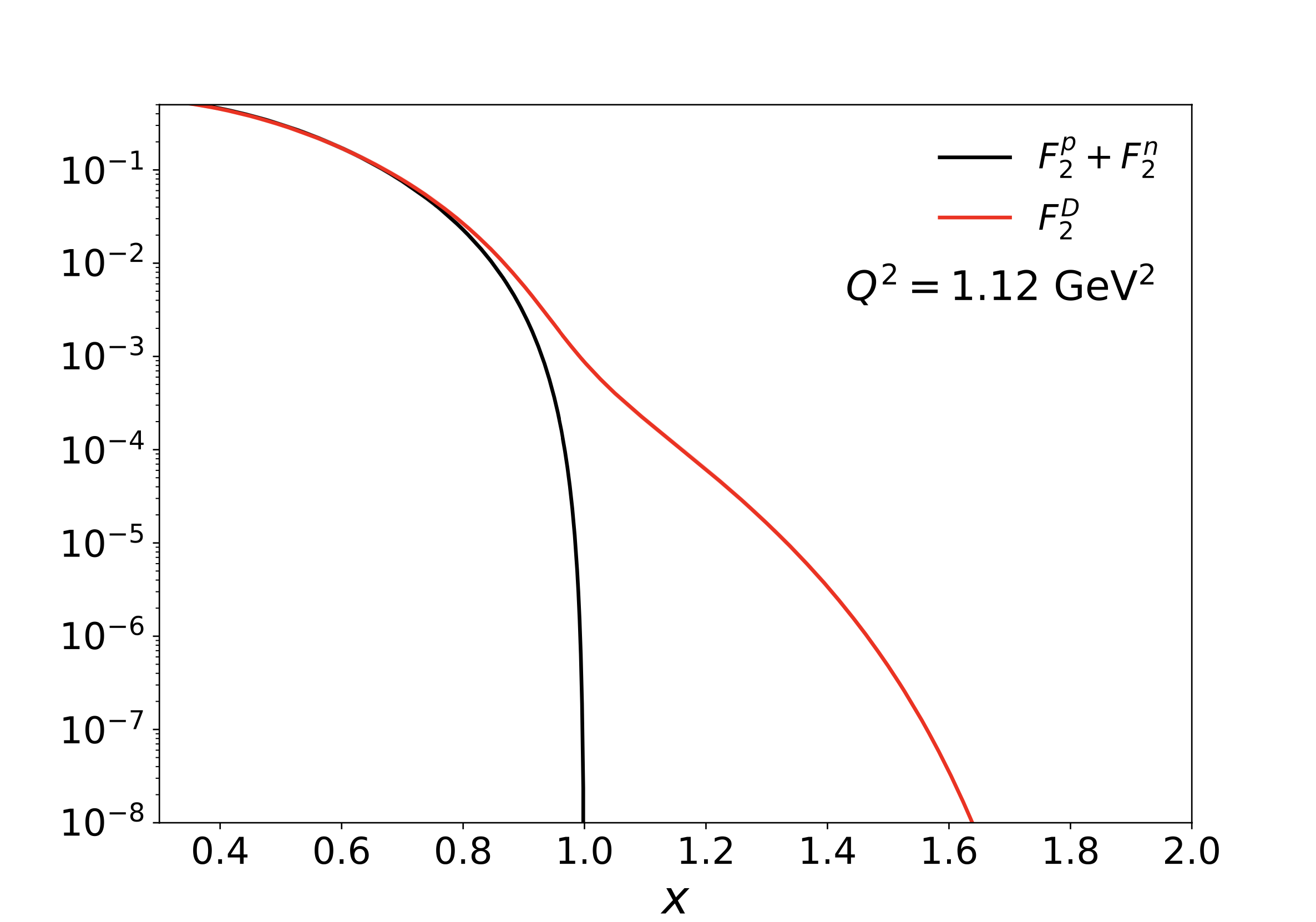}
    \caption{(color online) DIS $F_2$ structure functions on a logarithmic $y$-axis, evaluated at $Q^2 = 1.12$ GeV$^2$. The solid black line is the sum of the free proton and neutron structure functions and the solid red line was obtained by using Eq. (\ref{convolution_formula}).}
    \label{fig:deut_F2}
\end{figure}

Finally, using Eq. (\ref{convolution_formula}) we obtained $F_2^D$, evaluated at $Q^2 = 1.12$ GeV$^2$. The results are presented in Fig. \ref{fig:deut_F2}. The value used for the constant $\eta$ in Eq. (\ref{delta_r_ansatz}) was taken to be $0.4 \pm 0.1$, motivated from results in the nLFHQCD model \cite{Kim:2022lng}. Discussion on $\eta$ in the BLC-PLC model can be found in Appendix A.

\begin{figure}[H]
    \centering
    \includegraphics[width=6.5in]{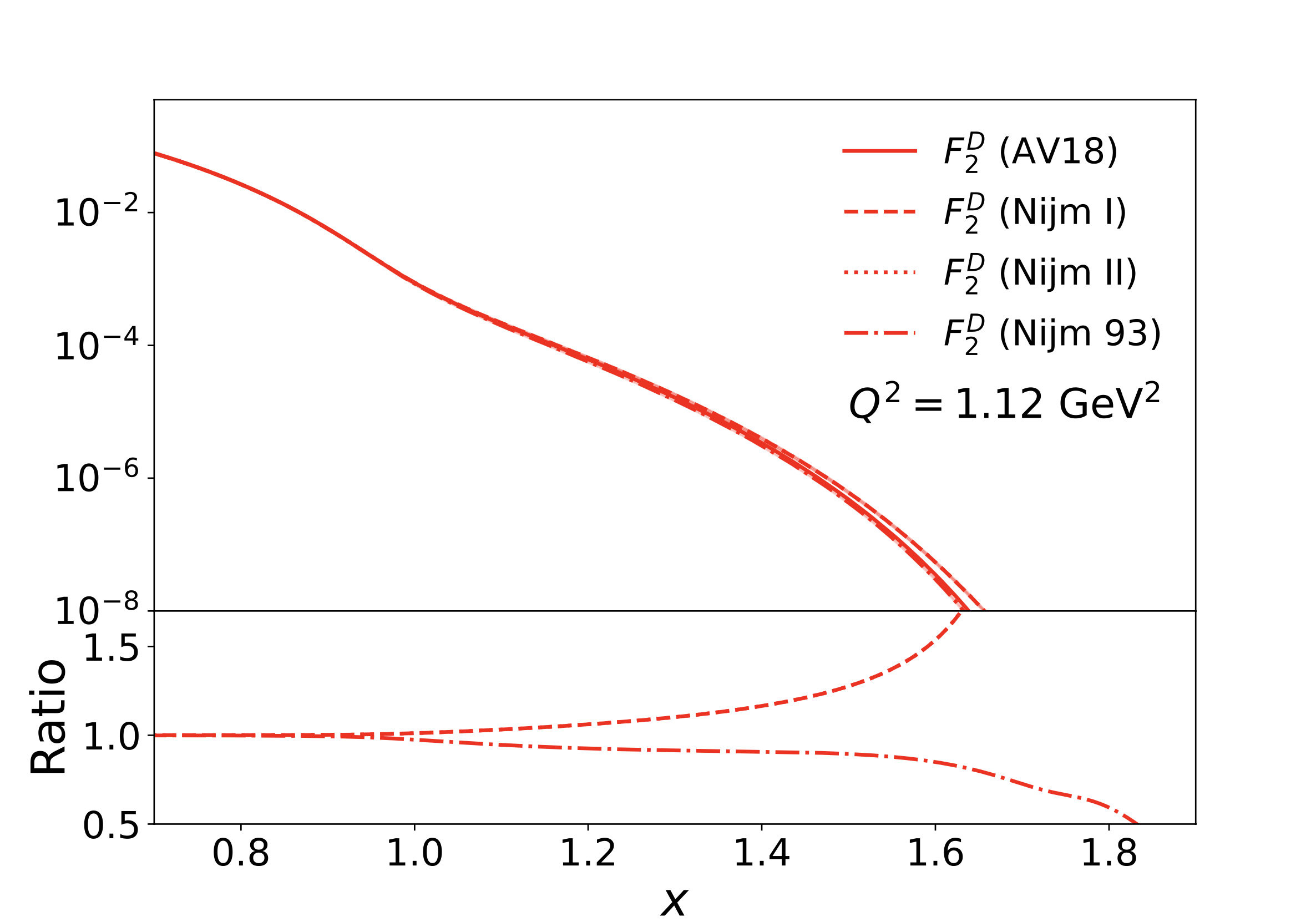}
    \caption{(color online) (top) Results for $F_2^D$, Eq. (\ref{convolution_formula}), using different deuteron wavefunctions. The sold red line uses the AV18 wavefunction and the dashed, dotted, and dot-dashed red lines use the NijmI, NijmII, and Nijm93 wavefunctions respectively. (bottom) The ratios of the Nijm $F_2^D$ results with respect to the AV18 results. Note that NijmI and NijmII lines are overlapped.}
    \label{fig:deut_F2_diff_wfk}
\end{figure}

Fig. \ref{fig:deut_F2_diff_wfk} presents results for $F_2^D$, Eq. (\ref{convolution_formula}), using the Nijmegen (NijmI, NijmII, Nijm93) \cite{Stoks:1994wp,deSwart:1995mb} and AV18 wavefunctions \cite{Wiringa:2013ala}, displaying the model's sensitivity to different nucleon-nucleon potentials. All deuteron wavefunctions were normalized according to Eq. (\ref{normalization}). We find that results using different nucleon-nucleon potentials agree in the $x < 1$ region, and begin to diverge when $x > 1$, with differences as large at 50\%. 

\begin{figure}[H]
    \centering
    \includegraphics[width = 6.5in]{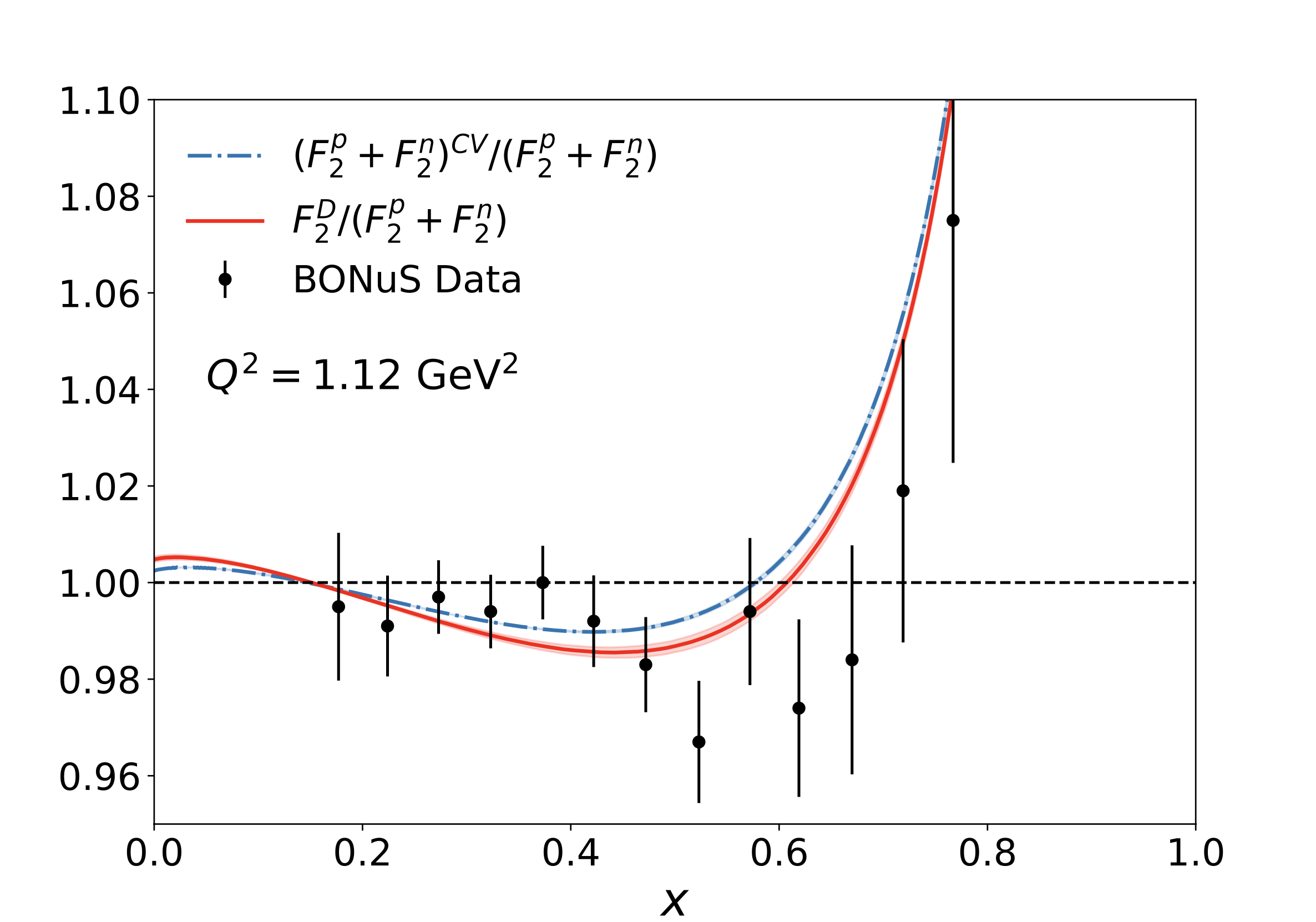}
    \caption{(color online) All DIS $F_2$ quantities are evaluated at $Q^2 = 1.12$ GeV$^2$. The solid red line is the convolution model nLFHQCD result using Eq. (\ref{convolution_formula}), the dot-dashed blue line was obtained by using free nucleon structure functions in Eq. (\ref{convolution_formula}), where the superscript $CV$ means 'convolution', and the filled black points are experimental results obtained from the BONuS experiment \cite{Fenker:2008zz,CLAS:2011qvj,CLAS:2014jvt,Griffioen:2015hxa}.}
    \label{fig:deut_emc}
\end{figure}

Fig. \ref{fig:deut_emc} presents the results for the EMC ratio for deuterium. All DIS $F_2$ quantities are evaluated at $Q^2 = 1.12$ GeV$^2$. Our results for $F_2^D/(F_2^p + F_2^n)$ give a $\chi^2 = 1.03$, in very good agreement with data obtained from the BONuS experiment, which extracted $F_2^n/F_2^D$ by using a spectator tagging technique on semi-inclusive electron-deuteron collisions \cite{Fenker:2008zz,CLAS:2011qvj,CLAS:2014jvt,Griffioen:2015hxa}. Notice that simply applying Fermi smearing to $F_2^p + F_2^n$ yields a deuterium EMC effect that captures data as well, with a $\chi^2 = 1.18$.


\section{Addition of Six-Quark Cluster}
With the light-cone convolution model $F_2^D$ result, we can now explore ideas outside conventional nuclear physics to include off-shell effects in the superfast region. Motivated towards understanding intersections between nuclear physics and QCD, we model the PDFs for an exotic six-quark state to make predictions on its contribution to $F_2^D$. Deuterium can occupy a six-quark state through quantum fluctuations, causing the proton and neutron to overlap completely. This six-quark cluster/bag, compared to a bound proton and neutron, allows for a greater sharing of momentum between the quarks in deuterium, enhancing the distribution of high-momentum quarks \cite{arrington:2006pr}. However, since most of deuteron's properties can be described by the picture of a bound proton and neutron with pionic effects, we expect the six-quark bag probability, $P_{6q}$, to be very small. A previous study investigated the contribution of a six-quark state to the $b_1$ structure function of the deuteron \cite{Miller:2013hla}. The study used a six-quark probability of $P_{6q} = 0.15$\% to match the experimental extraction of $b_1$ at $x = 0.452$ by the HERMES Collaboration \cite{HERMES:2005pon}. Since this value for the probability was obtained from fitting to one point, we will use it as a conservative upper-bound for $P_{6q}$. \\

Now we need to model the PDF of a six-quark hadronic state in order to get its DIS $F_2$ structure function,

\begin{equation} \label{F2_6q}
    F_2^{6q}(x,Q^2) = \frac{4}{9} \, \frac{x}{2} \, u^{6q}(x/2,Q^2) + \frac{1}{9} \, \frac{x}{2} \, d^{6q}(x/2,Q^2).
\end{equation}

\noindent Where $u^{6q}(x/2,Q^2)$ and $d^{6q}(x/2,Q^2)$ are the up and down PDFs of the six-quark state. To include the contributions of a six-quark cluster to $F_2^D$, we added Eq. (\ref{F2_6q}) to Eq. (\ref{convolution_formula}), multiplying both terms by a six-quark probability factor \textemdash \, not applying the convolution model to Eq. (\ref{F2_6q}) as a six-quark cluster does not have moving-nucleonic components. 

\begin{equation}\label{conv_6q}
    F_2^D(x,Q^2) = (1.0 - P_{6q}) \sum_N \int_{x}^{A} \frac{d\alpha}{\alpha} \int d\boldsymbol{k}_{\perp} \,  \rho_{N/D}(\alpha,\boldsymbol{k}_{\perp}) \, \tilde{F}_2^N(x/\alpha, \alpha, \boldsymbol{k}_{\perp}, Q^2) \, + P_{6q} F_2^{6q}(x,Q^2).
\end{equation}

\noindent Now, the theoretical issue is to determine an ansatz for the PDFs of a six-quark state in deuterium. 

\subsection{LFHQCD Six-Quark Ansatz}
Ref. \cite{Gutsche:2016lrz} utilized the LFHQCD framework to develop expressions for deuteron form factors. They accomplished this by working with an effective action that comprised of vector fields dual to the deuteron, and modifying the strength of the effective potential in the wave equation to fit experimental data. Although the results are promising, we believe this to be outside LFHQCD's realm of applicability, as the framework is constructed to study individual hadrons, not a collective nucleus. This idea is supported as Ref. \cite{Gutsche:2016lrz} had to modify the effective potential by roughly 100\% to fit their results to experimental data, which is not physically motivated. Instead, we believe their results can be used to predict the PDFs of a six-quark hadronic component in the deuteron. \\

We modified the results in Ref. \cite{Gutsche:2016lrz} by changing the strength of the effective potential back to its universal quantity, $\sqrt{\lambda} = 0.548$ GeV, which incorporates the correct Regge trajectories for mesons and hadrons, and by shifting the vector mesons mass poles to their physical twist-2 locations in the bulk to boundary photon propagator \cite{Brodsky:2014yha}. Using the procedure to obtain valence PDFs from elastic form factors outlined in Ref. \cite{deTeramond:2018ecg}, we found that six-quark PDFs use the same $q_\tau$ as in Eq. (\ref{q_tau}), which incorporates Regge behavior at small $x$, and inclusive counting rules at $x \rightarrow 1$: 

\begin{equation} \label{u_d_6q}
    u^{6q}(x/2,Q^2_o) = d^{6q}(x/2,Q^2_o) = \frac{3}{2} \, q_6(x/2,Q^2_o).
\end{equation}

\begin{figure}[H]
    \centering
    \includegraphics[width=6.5in]{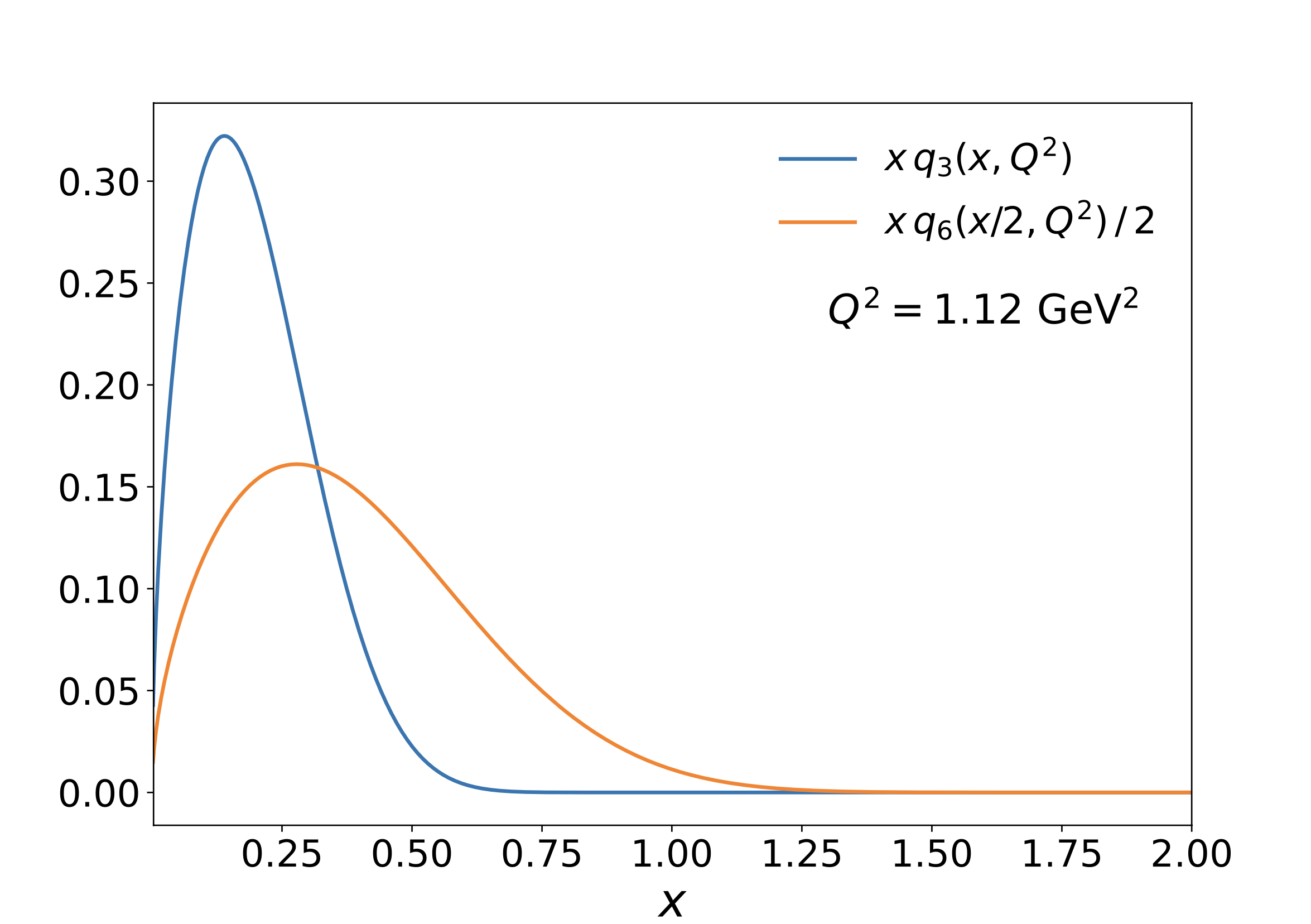}
    \caption{Comparison between $xq_3(x,Q^2)$ distribution used in bound-nucleon PDFs in Eq. (\ref{mod_p_u}, \ref{mod_p_d}), and the six-quark PDF ansatz used in Eq. (\ref{u_d_6q}), $xq_6(x/2,Q^2)/2$, evaluated at $Q^2 = 1.12$ GeV$^2$.}
    \label{fig:q_6_dist}
\end{figure}

\noindent Fig. \ref{fig:q_6_dist} shows a comparison between $xq_3(x,Q^2)$, used in the bound-nucleon PDFs, and the exotic six-quark PDF ansatz, $xq_6(x/2,Q^2)/2$.

\noindent By combining Eqs. (\ref{F2_6q}, \ref{u_d_6q}) we obtained:

\begin{equation} \label{F2_6q_lfhqcd}
    F_2^{6q}(x,Q^2_o) = \frac{5}{9} \, \frac{x}{2} \, \left( \frac{3}{2} \, q_6(x/2,Q^2_o) \right).
\end{equation}

\noindent Notice that the argument of $q_6$ is $x/2$, this ensures that the up and down six-quark PDFs extend to $x = 2$.

\begin{figure}[H]
    \centering
    \includegraphics[width=6.5in]{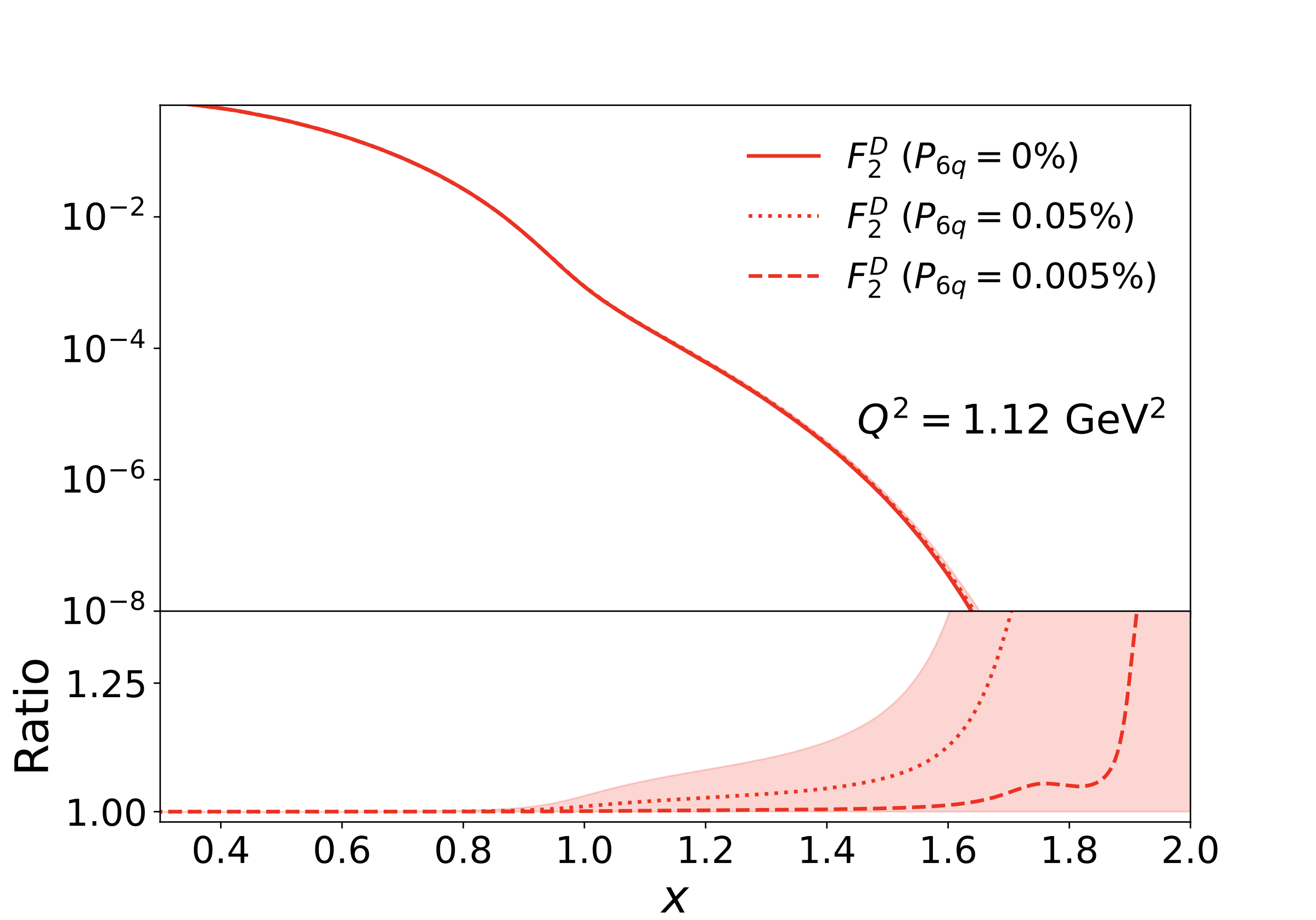}
    \caption{(color online) (top) DIS $F_2$ structure functions on a logarithmic $y$-axis. All $F_2$ quantities are evaluated at $Q^2 = 1.12$ GeV$^2$. (bottom) The ratios $F_2^D(P_{6q}) / F_2^D(P_{6q} = 0\%)$. The predicted six-quark contribution is displayed as a filled-in red volume, but is difficult to discern on the top plot with the given axis scaling. The lower boundary of the volume is $P_{6q} = 0 \%$ and the upper boundary is $P_{6q} = 0.15\%$. The red dotted and dashed lines display $P_{6q} = 0.05\%$ and $P_{6q} = 0.005\%$ respectively, and are shown to clarify the trend of $F_2^D$ with varying $F_2^{6q}$ contributions.}
    \label{fig:F2_6q}
\end{figure}

Fig. \ref{fig:F2_6q} presents the predictions of the LFHQCD six-quark model to our convolution model results from Section II, evaluated at $Q^2 = 1.12$ GeV$^2$.

\begin{figure}[H]
    \centering
    \includegraphics[width=6.5in]{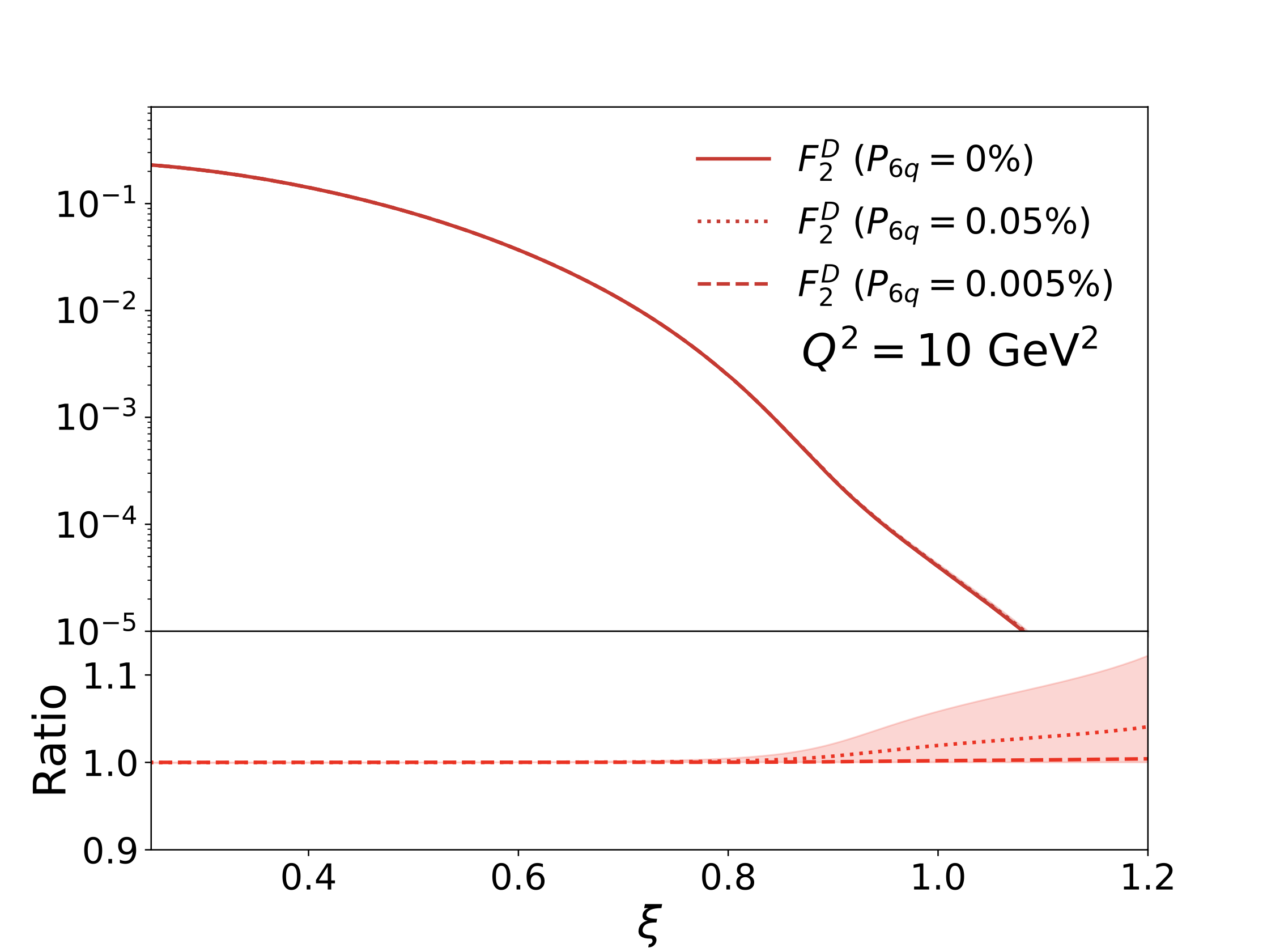}
    \caption{(color online) (top) DIS $F_2$ structure functions on a logarithmic $y$-axis, as a function of the Nachtmann variable, $\xi$, evaluated at $Q^2 = 10$ GeV$^2$. (bottom) The ratios $F_2^D(P_{6q}) / F_2^D(P_{6q} = 0\%)$. The predicted six-quark contribution is displayed as a filled-in red volume, but is difficult to discern on the top plot with the given axis scaling. The lower boundary of the volume is $P_{6q} = 0 \%$ and the upper boundary is $P_{6q} = 0.15\%$. The red dotted and dashed lines display $P_{6q} = 0.05\%$ and $P_{6q} = 0.005\%$ respectively, and are shown to clarify the trend of $F_2^D$ with varying $F_2^{6q}$ contributions.}
    \label{fig:F2_6q_nacht_evol}
\end{figure}

\noindent Fig. \ref{fig:F2_6q_nacht_evol} presents the same information as Fig. \ref{F2_6q}, but as a function of the Nachtmann variable, $\xi = 2x/(1 + \sqrt{1 + 4m^2x^2/Q^2})$, a common prescription for target mass corrections and displays scaling even at large values of $\xi$ (for more discussion see Refs. \cite{Nachtmann:1973mr,arrington:2006pr,Fomin:2010ei}). The results are evaluated at $Q^2 = 10$ GeV$^2$, kinematics that are within the proposed reach of 12 GeV JLab experiments \cite{arrington:2006pr, Hen:2014vua}). To accomplish this, the PDFs are evolved to a higher scale with the Dokshitzer-Gribov-Lipatov-Altarelli-Parisi (DGLAP) equations \cite{Altarelli:1977zs,Dokshitzer:1977sg,Gribov:1971zn} using the APFEL package \cite{Bertone:2013vaa}. The inputs for DGLAP evolution, such as the initial scale, renormalization scheme, and heavy quark thresholds, are identical to the ones used to study PDFs in LFHQCD in Ref. \cite{deTeramond:2018ecg}. 

\section{Discussion/Conclusion}
The results presented in this paper are an extension to our previous study in Ref. \cite{Kim:2022lng}. Focusing on the deuteron, we applied Fermi motion effects, using a variety of different wavefunctions, to extend the nLFHQCD model beyond $x > 1$. In doing so, we found that our model is in excellent agreement to BONuS data for the EMC ratio \cite{Fenker:2008zz,CLAS:2011qvj,CLAS:2014jvt,Griffioen:2015hxa}, and that the effects of different nucleon-nucleon potentials become significant in the superfast region. On top of conventional nuclear physics, we implemented the contributions of an exotic six-quark state to $F_2^D$ using a LFHQCD ansatz \cite{Gutsche:2016lrz}. We found that the six-quark distribution enhances the $x > 1$ region, while minimally affecting $x < 1$ \textemdash \, displaying correct qualitative behavior as a six-quark cluster allows for a greater sharing of momentum between quarks, enhancing the high-momentum behavior of $F_2^D$, while minimally affecting the low-momentum region. We displayed the predictions of the six-quark ansatz to $F_2^D$ for $0 < P_{6q} < 0.15\%$ in Fig. \ref{fig:F2_6q}, with the upper-bound in $P_{6q}$ motivated by Ref. \cite{Miller:2013hla}. We found that a small six-quark probability, $P_{6q} = 0.15\%$ can lead to large enhancements in the superfast region, enhancements greater than $25\%$ for $x > 1.6$. Furthermore, we found that the effects of different nucleon-nucleon potentials are around the same magnitude as six-quark effects in our model. With the proposed $12$ GeV experiments at JLab, we hope to test these predictions against experimental data in the near future. Deviations from the following predictions could be an indication of more interesting physics in the superfast region. 

\section{Acknowledgements}
D. N. Kim and G. A. Miller would like to thank John Arrington for motivating this study and Adam Freese, Mark Strikman, Stanley J. Brodsky, and Guy F. de Teramond for useful discussions. This work was supported by the U. S. Department of Energy Office of Science, Office of Nuclear Physics under Award Number DE-FG02-97ER-41014.

\bibliography{references}{}

\section*{Appendix A: Relationship between Light-Cone and Feynman Virtuality in BLC-PLC Model}
Firstly, we need to determine the connection between Eq. (\ref{virtuality}) and the nucleon-nucleon potential. Using light-cone perturbation theory, one can obtain the Weinberg Equation, which in the vicinity of the deuteron bound state, becomes the following Schrodinger-like equation for the wavefunction of the deuteron \cite{Weinberg:1966jm,Frankfurt:1981mk}:

\begin{equation}\label{weinberg_eq}
    (M_{1,2}^2 - M_D^2) \, \psi_D(\alpha,\boldsymbol{k}_{\perp}) = \Gamma(\alpha,\boldsymbol{k}_{\perp}) = \int V(\alpha ', \boldsymbol{k}_{\perp} ', \alpha,\boldsymbol{k}_{\perp}) 
    \psi_D(\alpha ',\boldsymbol{k}_{\perp} ') \frac{d\alpha '}{\alpha ' (2 - \alpha ')} \frac{d^2\boldsymbol{k}_{\perp}'}{(2\pi)^3},
\end{equation}

\begin{equation}
    M_{1,2}^2 = 4 \, \frac{m^2 + \boldsymbol{k}_{\perp}^2}{\alpha (2 - \alpha)}.
\end{equation}

\noindent Here $M_{1,2}^2$ is the invariant mass of the two nucleon system, $M_D^2$ is the squared mass of the deuteron, $V(\alpha ', \boldsymbol{k}_{\perp} ', \alpha,\boldsymbol{k}_{\perp})$ is the nucleon-nucleon potential, and $\psi_D(\alpha,\boldsymbol{k}_{\perp})$ is the light-cone deuteron wavefunction. The symbol $\Gamma(\alpha,\boldsymbol{k}_{\perp})$ is also known as the bound state vertex function. Neglecting all but nucleonic degrees of freedom, and identifying the $pn$ components of the deuteron wavefunction, we can use Eq. (\ref{pn_momentum}) to relate the light-cone deuteron wave equation, Eq. (\ref{weinberg_eq}), to the conventional non-relativistic Schrodinger equation for the deuteron:

\begin{equation}\label{nonrel_wf_eq}
    4 \, (k^2 + k_D^2) \, \psi_D(k) = \Gamma(k) = \int V(k,k') \, \psi_D(k')\,  \frac{d^3 \boldsymbol{k}'}{(2\pi)^3 \sqrt{m^2 + k'^2}}.
\end{equation}

\noindent Where $k_D^2 = m^2 - (M_D^2 \, / \, 4)$. Using Eq. (\ref{virtuality}), we can express the left-hand side of Eq. (\ref{nonrel_wf_eq}) as:

\begin{equation}\label{virt_relation}
    4 \, (k^2 + k_D^2) \, \psi_D(k) = - M_D^2 \, \mathcal{V}_{pn/D}(k) \, \psi_D(k).
\end{equation}

\noindent With Eqs. (\ref{nonrel_wf_eq}, \ref{virt_relation}), we get:

\begin{equation}\label{lc_virt_experssion}
     \mathcal{V}_{pn/D}(\alpha,\boldsymbol{k}_{\perp}) = - \frac{1}{M_D^2} \frac{\Gamma(k)}{\psi_D(k)},
\end{equation}

\noindent displaying the relationship between light-cone virtuality for the deuteron, the deuteron wavefunction, and the nucleon-nucleon potential. \\

Now we want to apply the BLC-PLC model used in Ref. \cite{Kim:2022lng}, which treats the nuclear potential with a number, to Eq. (\ref{lc_virt_experssion}). The simplified nucleon-nucleon potential in momentum space must take the following form to match the potential used in the BLC-PLC model:

\begin{equation}\label{nn_potential_momentum_space}
    V(k,k') = (2m)^2 \, |U_{NN}| \, (2\pi)^3 \, \delta^{(3)}(\boldsymbol{k} - \boldsymbol{k} ').
\end{equation}

\noindent Where $|U_{NN}|$ is the simplified nucleon-nucleon potential used for deuteron in the BLC-PLC model in Ref. \cite{Kim:2022lng}, and the $(2m)^2$ factor is the relativistic normalization factor which connects relativistic and non-relativistic scattering matrix elements. Plugging in Eq. (\ref{nn_potential_momentum_space}) into Eq. (\ref{lc_virt_experssion}) we obtain:

\begin{equation}
    \mathcal{V}_{pn/D}(\alpha,\boldsymbol{k}_{\perp}) = - \frac{(2m)^2}{M_D^2} \frac{|U_{NN}|}{\sqrt{m^2 + k^2}},
\end{equation}

\noindent Which in the non-relativistic limit, and taking $2m \approx M_D$, becomes:

\begin{equation}\label{BLC_PLC_LC_virt}
    \mathcal{V}_{pn/D}(\alpha,\boldsymbol{k}_{\perp}) \approx - \frac{|U_{NN}|}{m}.
\end{equation}

\noindent Ref. \cite{Kim:2022lng} uses $\mathcal{V} = (m^2 + k^2)/m^2$ and obtained the following expression for virtuality in the BLC-PLC model:

\begin{equation}
    \mathcal{V} = - \frac{2 \, |U|}{m}.
\end{equation}

\noindent Thus we find that light-cone virtuality is exactly half the Feynman virtuality in the BLC-PLC model. 
Defining the average virtuality as:

\begin{equation}
    <\mathcal{V}_{np/D}> \, = \int_0^A \frac{d \alpha}{\alpha} \int d^2\boldsymbol{k}_{\perp} \mathcal{V}_{np/D}(\alpha,\boldsymbol{k}_{\perp}) \, \rho_{np/D}(\alpha,\boldsymbol{k}_{\perp}),
\end{equation}

\noindent we obtain $<\mathcal{V}_{np/D}> \, = -0.0235$, which corresponds to a value of $-0.047$ if using Feynman virtuality. This is in agreement with the average Feynman virtuality obtained in Ref. \cite{Wang:2020uhj}. With Eq. (\ref{BLC_PLC_LC_virt}) we determine, using results in Ref. \cite{Kim:2022lng},

\begin{equation}\label{delta_r_LC}
    \delta r_{pn/D}(\alpha,\boldsymbol{k}_{\perp}) = - \frac{m}{4} \frac{\mathcal{V}_{np/D}(k)}{\bar{\Delta}}.
\end{equation}

\noindent Relating Eq. (\ref{delta_r_LC}) with Eq. (\ref{delta_r_ansatz}), we find that the constant $\eta = m/(4\bar{\Delta})$. In our study, we used $\eta = 0.4 \pm 0.1$ which gives $\bar{\Delta} \approx 587 \pm 147$ GeV, in agreement with the BLC-PLC model which limits $\bar{\Delta}$ to be greater than or equal to the difference between the Roper resonance and nucleon mass.

\end{document}